\documentclass[aps,prd,superscriptaddress,showkeys,showpacs,floatfix,eqsecnum,twocolumn]{revtex4}

\usepackage{graphics}
\usepackage{graphicx}
\usepackage{dcolumn}
\usepackage{bm}
\usepackage{amssymb,amsmath,graphicx,graphics,color,epstopdf}
\usepackage{hyperref}
\hypersetup{
    colorlinks=true,
    linkcolor=blue,
    filecolor=magenta, citecolor=blue,    urlcolor=blue,
}

\begin{document}


\title{Hawking Radiation via Gauss-Bonnet Theorem}

\author{A. \"{O}vg\"{u}n}
\email{ali.ovgun@pucv.cl}
\affiliation{Instituto de F\'{\i}sica, Pontificia Universidad Cat\'olica de Valpara\'{\i}%
so, Casilla 4950, Valpara\'{\i}so, Chile.}
\affiliation{Physics Department, Faculty of Arts and Sciences, Eastern Mediterranean
University, Famagusta, North Cyprus, via Mersin 10, Turkey.}

\author{\.{I}. Sakall{\i}}
\email{izzet.sakalli@emu.edu.tr}
\affiliation{Physics Department, Faculty of Arts and Sciences, Eastern Mediterranean
University, Famagusta, North Cyprus, via Mersin 10, Turkey.}

\date{\today}
\begin{abstract}

In this paper, we apply the topological method to the various black holes to derive their Hawking temperature. We show that the this method can easily be employed to compute the Hawking temperature of black holes having spherically symmetric topology. Therefore, we conclude that the topological method provides a consistent formula to achieve the Hawking temperature. 

\end{abstract}


\maketitle

\section{Introduction}

Before Hawking, the black holes were thought to be just like the vacuum cleaner that sucked everything inside the universe. Black holes are very intense objects, so even light cannot escape from it. Fortunately, Hawking came an idea that \cite{Hawking:1974sw} something does escape a black hole: radiation. This became the one of the most important discovery of Hawking and the equation is named for him: Hawking radiation. His works \cite{Hawking:1974sw,Hawking:1975sw} merged two different theories such as quantum mechanics and general relativity. Afterwards, various methods (\cite{Gibbons:1977mu,Unruh:1976db,Damour:1976jd,Parikh:1999mf,Srinivasan:1998ty,Akhmedov:2006pg,Akhmedov:2008ru,Gibbons:1976ue}) have been proposed to obtain the Hawking radiation. For example, Hawking and Gibbons derived the Hawking temperature using the the Euclidean path integral for the gravitational field \cite{Gibbons:1976ue}. Then, Wilczek and Parikh showed that it is possible to derive the Hawking radiation via a quantum tunnelling process \cite{Parikh:1999mf}. This phenomenon gained a lot of attention and has been applied on different compact objects such as black holes, wormholes, black strings, and cosmic strings (see for example \cite{Kruglov:2014jya,Sakalli:2015jaa,Sakalli:2015taa,Sakalli:2014sea,Sakalli:2015mka,Sakalli:2016cbo,Kuang:2017sqa,Becar:2010zza, Singh:2019oxh,Singh:2019wtn,KenedyMeitei:2019cna,Singh:2017mqv,sucu,Ali:2007sh,HossainAli:2004xj,Li:2015oia,Sharif:2013nda,Kerner:2006vu,Sakalli:2016ojo,Sakalli:2012zy,Sakalli:2010yy,Pasaoglu:2009te,Sakalli:2016fif,Banerjee:2007qs,Banerjee:2008sn}).

Recently, Robson, Villari, and Biancalana have shown that Hawking temperature of a black hole can also be topologically derived \cite{Robson:2018con,Robson:2019yzx} such as obtaining the entropy of the black hole from topology \cite{Gibbons:1994ff,Liberati:1997sp,Olea:2005gb,Miskovic:2009bm}. The topological method is based on the Gauss-Bonnet theorem and the Euler characteristic, which are invariants of the topology. In the topological method, one uses the Euclidean geometry of the 2-dimensional spacetime (by applying the Wick rotation \cite{Gibbons:1977mu}) to derive Hawking temperature without losing any information of the 4-dimensional spacetime. In fact, the lower dimensional base is a playground for the holography and the stored information .

In this work, we shall apply the topological method to various spacetimes of black holes such as dyonic Reissner-Nordstr{\"o}m black hole, Schwarzschild-electromagnetic black hole, quantum corrected black hole, Schwarzschild-like black hole in a bumblebee gravity model and linear dilaton black hole. For this purpose, first we briefly review the topological method in the following section.

\section{Computation of Hawking temperature via Gauss-Bonnet
Theorem}\label{sec:temp}

The topological method computes the black hole temperature in a simple way via the 2-dimensional Euler characteristic and the Gauss-Bonnet theorem \cite{Robson:2018con,Robson:2019yzx}. To understand the formalism of this method, let us first consider the following 4-dimensional spherically symmetric and static black hole metric
\begin{equation}
 ds^2 = -f (r) d t^2 + \frac{1}{f (r)} dr^2+ r^2 d \theta^2 + r^2 sin^2 \theta d \phi^2 , \label{metric}
\end{equation}

which recasts in the following 2-dimensional Euclidean Schwarzschild coordinates via the Wick rotation ($\tau = it$) and $\theta=\frac{\pi}{2}$:

\begin{equation}
 ds^2 = f (r) d \tau^2 + \frac{1}{f (r)} dr^2. \label{metric2}
\end{equation}

One can easily compute the Ricci scalar of the above metric as
\begin{equation}
R=-{\frac {{\rm d}^{2}}{{\rm d}{r}^{2}}}f \left( r \right) \label{R}.\end{equation}

The topological formula for the Hawking temperature of the above two-dimensional black holes is given by
\begin{equation}
 T_{H} = \frac{\hbar c}{4 \pi \chi k_{B}} 
  \sum_{j \leq \chi} \int_{r_{h_{j}}} \sqrt{g} R dr, \label{for}
\end{equation}
where $\hbar$ is the Planck constant, $c$ denotes the speed of light, and the Boltzmann's constant is represented by $k_{\mathrm{B}}$. Moreover, Ricci scalar $R$ is only the function of the `spatial' variable $r$, $g$ is the Euclidean metric determinant, $r_h$ stands for the horizon  with $j$-th
Killing horizon, and $\chi$ denotes the Euler characteristic of 
Euclidean geometry. Besides, $\sum_{j \leq \chi}$  shows the summation over the
Killing horizons. From now on, without any loss of generality, we set the all physical constants to unity to simplify the formulae and metrics. 

\section{Hawking Temperature of Dyonic Reissner Nordstr{\"o}m Black Hole}

In this section, we shall apply the method defined in previous section to the 4-dimenional dyonic Reissner Nordstr{\"o}m black hole (DRNBH) \cite{Sakalli:2016ojo,DRNBH2,DRNBH3}. To this end, we shall first consider the 2-dimensional DRNBH:

\begin{equation}
  ds^2 = -f(r) dt^2 +
  \frac{1}{f(r)} dr^2 , \label{DR}
\end{equation}

where 
\begin{equation}
f ( r ) = 1 - \frac { 2 M } { r } + \frac { Q ^ { 2 } + P ^ { 2 } } { r ^ { 2 } },
\end{equation}
which admits the following event horizon:
\begin{equation}
r _ { h } = M \pm \sqrt { M ^ { 2 } - Q ^ { 2 } - P ^ { 2 } }.
\end{equation}

By using the Wick rotation, we turn the Riemannian metric \eqref{DR} of DRNBH to the Euclidean spacetime:
\begin{eqnarray}
  ds^2=\left( 1 - \frac { 2 M } { r } + \frac { Q ^ { 2 } + P ^ { 2 } } { r ^ { 2 } }\right) d \tau^2 \\ \notag
  +\frac{1}{\left( 1 - \frac { 2 M } { r } + \frac { Q ^ { 2 } + P ^ { 2 } } { r ^ { 2 } } \right)} dr^2 .
\end{eqnarray}

By following \cite{Robson:2018con} and knowing that 2-dimensional Euclidean spacetimes have topology of $R^2$ and Euler characteristic $\chi = 1$  \cite{Gibbons:1994ff}, we utilize the formula (\ref{for}) to get the Hawking temperature of the DRNBH. So, we have 
\begin{equation}
 T_{H} = \frac{1}{4 \pi} 
  \int_{r_h} \sqrt{g} R dr, \label{my1}
\end{equation}
in which $g=1$ and the Ricci scalar of the DRNBH:
\begin{equation}
R=\,{\frac {4M}{{r}^{3}}}-\,{\frac {6({P}^{2}+{Q}^{2})}{{r}^{4}}}.
\end{equation}

After evaluating the integral \eqref{my1}, we get the temperature of the DRNBH as:
\begin{equation}
  T_{\mathrm{H}} = \frac { \sqrt { M ^ { 2 } - Q^2-P^2 } } { 2 \pi \left( M + \sqrt { M ^ { 2 } - Q^2-P^2 } \right) ^ { 2 }, }\end{equation}

which is fully in agreement with the standard Hawking temperature result ($T_{H}=\frac{\kappa}{2\pi}$) of the DRNBH \cite{Sakalli:2016ojo}.

\section{Hawking Temperature of Schwarzschild-Electromagnetic Black hole}

For the line-element \eqref{DR}, the metric function for the Schwarzschild-electromagnetic black hole \cite{Ovgun:2015jna} reads
\begin{equation}
    { f = 1 - \frac { 2 M } { r } + \frac { M ^ { 2 } \left( 1 - a ^ { 2 } \right) } { r ^ { 2 } } }. \end{equation} 
 Schwarzschild-electromagnetic black hole possesses a Killing horizon related to the time-isometry and its horizon is located at
\begin{equation}
r _ { h } = M ( 1 + a ).
 \end{equation}

Using the topological formula \eqref{my1} with the Ricci scalar of the Schwarzschild-electromagnetic black hole:
\begin{equation}
R=4\,{\frac {M}{{r}^{3}}}-6\,{\frac {{M}^{2} \left( -{a}^{2}+1 \right) 
}{{r}^{4}}},
\end{equation}
We get the temperature of the Schwarzschild-electromagnetic black hole from the topology as follows:
\begin{equation}
T_H= \frac { 1 } { 4 \pi } \frac { 2 a } { M ( a + 1 ) ^ { 2 } },
\end{equation}
which is nothing but its Hawking temperature \cite{SchwEM}.
\section{Hawking Temperature of Reissner Nordstr{\"o}m-like Black Hole with Quantum
Potential}
The metric function of the Reissner Nordstr{\"o}m-like black Hole with quantum
potential for the Euclidean spacetime \eqref{DR} is given by \cite{Ali:2015tva}:
\begin{equation}
f ( r ) = 1 - \frac { 2 M } { r } + \frac { \hbar \eta } { r ^ { 2 }}, \end{equation} 
where $\eta$ is a constant. The event horizon of this black hole is located at

\begin{equation}
r _ { h } = M \pm \sqrt { M ^ { 2 } - \eta \hbar },
\end{equation} 

and its Ricci scalar yields
\begin{equation}
R=4\,{\frac {M}{{r}^{3}}}-6\,{\frac {h\eta}{{r}^{4}}}. \label{RQ}
\end{equation}

After substituting Eq. \eqref{RQ} into the integral \eqref{my1} and in sequel making a straightforward computation, one gets the temperature of the black hole as follows
\begin{equation}
T _ { H } = \frac { \sqrt { M ^ { 2 } - \eta \hbar } } { 2 \pi \left( \sqrt { M ^ { 2 } - \eta \hbar } + M \right) ^ { 2 } }. \label{QBH}
\end{equation}
The result obtained in Eq. \eqref{QBH} perfectly fits with the value of the Hawking temperature given in \cite{Ali:2015tva}.

\section{Hawking temperature of Schwarzschild-like black hole in bumblebee gravity model}

The asymmetric metric functions of Schwarzschild-like black hole in the bumblebee gravity model (i.e., an exact vacuum solution from the gravity sector contained in the minimal standard-model
extension)  are given by \cite{Casana:2017jkc}:
\begin{equation}
g_{tt} =1 - \frac { 2 M } { r }, \end{equation}

and 
\begin{equation}
g_{rr}=( 1 + \ell ) \left( 1 - \frac { 2 M } { r } \right) ^ { - 1 }
\end{equation}
where $l$ is a constant and known as the Lorentz violating parameter \cite{Casana:2017jkc}. The event horizon of this black hole is nothing but the event horizon of the standard Schwarzschild black hole: $r_ { h } = 2M$ and it has the following Ricci scalar:
\begin{equation}
R=\,{\frac {4M}{{r}^{3} \left( 1+l \right) }}.
\end{equation}
Employing the formula of \eqref{my1}, we obtain the temperature of the Schwarzschild-like black hole in the bumblebee gravity model:
\begin{equation}
T _ { H } = \,{\frac {1}{8M\pi\,\sqrt {1+l}}},
\end{equation}
which is exactly equal to its standard Hawking temperature \cite{Casana:2017jkc}. 

\section{Hawking Temperature of linear dilaton black hole }
The 4-dimensional line-element of the static, spherically symmetric but  non-asymptotically flat linear dilaton black hole is given by \cite{Clement:2002mb,Clement:2007tw}
\begin{equation}
    \mathrm { d } s ^ { 2 } = - \frac { r - b } { r _ { 0 } } \mathrm { d } t ^ { 2 } + \frac { r _ { 0 } } { r - b } \mathrm { d } r ^ { 2 } + r _ { 0 } r \left( \mathrm { d } \theta ^ { 2 } + \sin ^ { 2 } \theta \mathrm { d } \phi ^ { 2 } \right), \label{ldbh}
\end{equation}
where $b=4M$ ($M$: quasilocal mass), $r_0=\sqrt{2}Q$ ($Q$: background electric charge), and the event horizon is $r_h=b$.
Since the topology of the linear dilaton spacetime \eqref{ldbh} is different than the Schwarzschild family of black holes. 
Thus, the formula for having the black hole temperature in 4-dimensional Euclidean spacetime becomes \cite{Robson:2019yzx}:

\begin{equation}
T _ { \mathrm { H } } = \frac { 1 } { 4 \pi  } \int d r \sqrt { g } \left( K _ { 1 } - 4 R _ { a b } R ^ { a b } + R ^ { 2 } \right) \label{74},
\end{equation}
where \begin{equation}\left( K _ { 1 } - 4 R _ { a b } R ^ { a b } + R ^ { 2 } \right)={\frac {b}{{r}^{3}{{\it r_{0}}}^{2}}} \label{75}. \end{equation}

Ricci scalar of the 4-dimensional linear dilaton black hole can be computed as follows
\begin{equation}
R=\left({\frac {r-b}{2{r}^{2}{\it r_0}}}\right). \label{ldbhRic}
\end{equation}

Then, substituting Eq. \eqref{75} into Eq. \eqref{74}, we obtain
\begin{equation}
T _ { H } =  \frac { 1 } { 4 \pi r _ { 0 } }. \label{finldbh}
\end{equation}
It is worth noting that the temperature of the linear dilaton black hole is independent of the mass parameter $b$. In fact, as can be seen from Eq. \eqref{finldbh}, the temperature depends only on constant $r_0$, which means that these black holes radiate isothermally \cite{Sakalli:2010yy,Pasaoglu:2009te}.

Alternatively, Charles W. Robson and Fabio Biancalana \cite{Robson:2018con, Robson:2019yzx} have very recently proven us (in accordance with the fruitful discussion between us) that one gets the same answer from the 2-dimensional Euclideanised spacetime from the following way. If we first divide the metric overall to $r/r_{0}$ and suppress the spherical part, we get

\begin{equation}
    \mathrm { d } s ^ { 2 } =  \frac { r(r - b) } { r _ { 0 }^2} \mathrm { d } t ^ { 2 } + \frac { r }  { (r - b) } \mathrm { d } r ^ { 2 }. \label{ldbh2}
\end{equation}
To find the temperature of linear dilaton black hole, we employ Eq. \eqref{for} and obtain
\begin{equation}
 T_{H} = \frac{1}{4 \pi \chi} 
  \int_{r_h} \sqrt{g} |R| dr=\frac { 1 } { 4 \pi r _ { 0 } }, \label{finito}
\end{equation}
in which $\sqrt{g}=\frac{r}{r_0}$ and the Ricci scalar of the Euclideanised linear dilaton black hole spacetime:
\begin{equation}
R=-\frac {b}{r^3}, \end{equation}

where the Euler characteristic used in Eq. \eqref{finito} is $\chi=1$.

\section{Conclusions}

In this paper, we have used the topological method, which considers the topological fractions together with the Gauss-Bonnet theorem. Then, we have applied the this method different spacetimes including the non-asymptotically flat ones. This new approach shows that Hawking radiation possesses a topological effect coming from the Euler characteristic of the spacetime. The Ricci scalar of the spacetime encodes all the information about the spacetime, so that it, with the Euler characteristic of the  metric, enables us to derive temperature of the black hole. 

To support the use of the topological method by samplings, we have derived the temperature of dyonic Reissner-Nordstr{\"o}m black hole, Schwarzschild-electromagnetic black hole, Reissner-Nordstr{\"o}m-like black hole with quantum potential, Schwarzschild-like black hole in a bumblebee gravity model, and  the linear dilaton black hole. Hence, we have verified the reliability of the method on different spacetimes. Hence, this new formula will break ground in this area.

The cylindrical spacetimes have different topology then the spherical ones. The application of the topological method on the black strings may reveal more information compared to the present case. This is the next stage of study that interests us.

\acknowledgments 
This work was supported by Comisi{\'o}n Nacional de Ciencias y Tecnolog{\'i}a of Chile through FONDECYT Grant $N^\mathrm{o}$
3170035 (A. {\"O}.).

\end{document}